\documentclass[conference]{IEEEtran}

\usepackage{graphicx}
\usepackage{subfigure}

\ifCLASSINFOpdf
\else
\fi
\usepackage{amsmath}
\usepackage{amsthm}
\begin{document}
%
\title{Minor probability events detection in big data: An integrated approach with Bayesian testing and MIM}
%
%
%

\author{Shuo~Wan, Jiaxun~Lu, Pingyi~Fan,~\IEEEmembership{Senior Member,~IEEE} and Khaled~B.~Letaief{*},~\IEEEmembership{Fellow,~IEEE}\\

\small
Tsinghua National Laboratory for Information Science and Technology(TNList),\\

Department of Electronic Engineering, Tsinghua University, Beijing, P.R. China\\
E-mail: wan-s17@mails.tsinghua.edu.cn, lujx14@mails.tsinghua.edu.cn, ~fpy@tsinghua.edu.cn\\
{*}Department of Electronic Engineering, Hong Kong University of Science and Technology, Hong Kong\\
Email: eekhaled@ece.ust.hk}

\maketitle

\begin{abstract}
The minor probability events detection is a crucial problem in Big data. Such events tend to include rarely occurring phenomenons which should be detected and monitored carefully. Given the prior probabilities of separate events and the conditional distributions of observations on the events, the Bayesian detection can be applied to estimate events behind the observations. It has been proved that Bayesian detection has the smallest overall testing error in average sense. However, when detecting an event with very small prior probability, the conditional Bayesian detection would result in high miss testing rate. To overcome such a problem, a modified detection approach is proposed based on Bayesian detection and message importance measure, which can reduce miss testing rate in conditions of detecting events with minor probability. The result can help to dig minor probability events in big data.
\end{abstract}


%
\IEEEpeerreviewmaketitle

\section{Introduction}
%
%
%
%
\IEEEPARstart{T}{he} rapid growth of data amount in the internet has triggered research on big data analysis. In some applications such as learning actions of customers, the common events which happen in high probabilities need to be dig out. However, in many applications such as surveillance of abnormal terms in big data, the important messages are often hidden in data with minor prior probabilities. Therefore, the precise detection of minor probability events is a crucial problem in big data analytics.

The two main crucial indices in detection problems are miss testing rate and false alarming rate. The miss testing rate represents the proportion of events of interest which are not detected by the employed approaches. The false alarming rate represents the proportion of irrelative events which are falsely detected as events of interest. In the literature, the commonly used detection algorithms are based on the Bayesian rule. When given the prior probabilities and the conditional probability distributions of the observations, it can detect an event with the observation data.

However, when the event of interest has a prior probability which is much smaller than others, the Bayes detector would result in high miss testing rate. Furthermore, the miss testing rate is actually the main concern of users in the minor probability events detection. Since such events do not happen frequently, miss testing could lose many important messages, which may cause wrong inference. However, the false alarming rate is considerably admitted as it only brings extra cost. In such applications, such extra cost is not cared as long as it can be constrained in an acceptable range. Therefore, the detector has to be improved to lower down the miss testing rate while remaining a reasonable false alarming rate.

The minor probability events detection has been considered in terms of computer vision by a lot of adapted works. In \cite{4407716}, the authors designed a novel algorithm for detection of certain types of unusual events using multiple fixed-location monitors. Each local monitor produces an alert based on its measurement and the alerts are integrated to make a final decision. Beforehand, the relative works also considered the problem of modelling and detecting abnormal actions in videos \cite{boiman2007detecting}. Among them, the tracking-based detection approaches \cite{johnson1996learning} and those using low-level information \cite{xiang2006beyond} are very common.

In the big data, some existing works mainly considered problems of outlier detection \cite{ramaswamy2000efficient}. In big data analysis, traditional theoretical measures such as Kolomogorov Complexity, Shannon entropy and relative entropy are still widely applied to describe the exceptional sets. Works such as \cite{lee2001information} adopted the combination of several different measures to characterize the abnormal sets for detection in big data. Moreover, with an information theoretic approach, the objective function related to factorization based on distribution was constructed to detect minority subset \cite{ando2006information}.

In this paper, the focused problem is to detect a certain minor probability event in big data. The prior probability of the event can be estimated from the experience. The probability distribution of observed data conditioned on the occurring event can also be obtained. Then a natural way is using the Bayesian detection. However, when the event of interest has a very small prior probability, the miss testing rate can be very high. To overcome this problem, the probability distribution is analyzed by a new term called the message importance measure (MIM).

In \cite{fan2016message}, a new measurement of the importance of message was proposed for detection of minor probability event. Then in \cite{7996803}, the parameter selection of the message importance measure was discussed. Based on the message importance measure, the divergence measure and storage code design were proposed in \cite{she2017amplifying}\cite{liu2017non}.
The message importance measure was proposed to target minority subsets in big data opposed to the conventional information theoretic measures. It focuses more on the significance of the anomaly events with small occurring probability in big data scenarios. By applying this new measure of the probability distribution, the minor probability events can be magnified. Then the miss testing rate for such events can be reduced in big data analysis.

The rest of the article is arranged as follows. In Section \uppercase\expandafter{\romannumeral2}, the considered problem of detecting minor probability events is stated. Besides the flaw of Bayesian method in this problem is introduced. In Section \uppercase\expandafter{\romannumeral3}, the definitions and properties of message importance measure (MIM) are reviewed. In Section \uppercase\expandafter{\romannumeral4}, the new MIM based detection method is proposed. Then in Section \uppercase\expandafter{\romannumeral5}, simulations to test the performance of the new developed detection approach are displayed.




\section{Problem Statement}
In big data analysis, there is a huge amount of data representing the observations of events. Supposing there is a set of events which may happen behind the data, the target is to determine the event of interest when the observation data comes up. Each event has a prior probability which can be estimated from the general experience. In this paper, the focus is to detect occurrence of the minor probability event from a large set of observation data. In this problem, the majority of the data is from the frequently occurring event, while only a small part of them is from the event of interest. Therefore, a high miss testing rate is unexpected. In this paper, the aim is to detect minor probability events from data sets with a low miss testing rate and an acceptable false alarming rate.

Considering a large probability event $A$ and a minor probability event $B$, their prior probabilities can be estimated as $w_{A}$ and $w_{B}$ which satisfy $w_{A} >> w_{B}$. There is a big data set with $N$ observed data denoted as $X=\{ x_{1}, x_{2}, ......, x_{N} \}$. For any of its component $x_{i}$ where $i \in \{ 1, 2, ......, N \}$, there is the conditional probability distribution $p(x_{i} | A)$ and $p(x_{i} | B)$. For simplicity, they are denoted as $p_{A}(x_{i})$ and $p_{B}(x_{i})$.

Under these circumstances, the aim is to determine the event occurring behind data $x_{i}$ in the big data set $X$. Traditionally, this is a Bayesian detection problem based on the prior probabilities. Under such strategies, the decision criterion should be
\begin{equation}
\{x_{i}|x_{i}\in A\}=\left \{ x_{i}|\frac{w_{A}p_{A}(x_{i})}{w_{B}p_{B}\left ( x_{i} \right )}>1 \right \}
\end{equation}
\begin{equation}
\{x_{i}|x_{i}\in B\}=\left \{ x_{i}|\frac{w_{B}p_{B}(x_{i})}{w_{A}p_{A}\left ( x_{i} \right )}>1 \right \}
\end{equation}

The criterions to evaluate such detection methods are miss testing rate $\alpha$ and false alarming rate $\beta$. Considering the prior probabilities of event $A$ and $B$, the overall error rate should be
\begin{equation}\label{Pe}
P_{e}=w_{B}\alpha + w_{A}\beta
\end{equation}
It has been proved that the Bayesian detection criterion is the optimal detection method which can achieve the smallest overall error $P_{e}$. The estimation of $P_{e}$ for Bayesian detection is given by the following Lemma 1.

\newtheorem{lemma}{Lemma}
\begin{lemma} (Chernoff)
Supposing the observation $X$ is conditioned on event $Q$, there are two assumptions. The prior probability of $Q=A$ is $w_{A}$ and the prior probability of $Q=B$ is $w_{B}$. In addition, the conditional distributions on the two events are separately $p_{A}(X)$ and $p_{B}(X)$. The miss testing rate is $\alpha$ and the false alarming rate is $\beta$. Then the overall error rate defined as (\ref{Pe}) should satisfy
\begin{equation}
-{\rm log(P_{e})} \rightarrow D^{*}
\end{equation}
where $D^{*}$ is the optimal index satisfying
\begin{equation}
D^{*}=D( p_{\lambda^{*}} || p_{A} )=D( p_{\lambda^{*}} || p_{B} )
\end{equation}
$D(.||.)$ is the K-L divergency and the distribution $p_{\lambda}$ is defined as
\begin{equation}
p_{\lambda}=\frac{p_{A}^{\lambda }(x)p_{B}^{1-\lambda }(x)}{\int p_{A}^{\lambda }(x)p_{B}^{1-\lambda }(x) dx }
\end{equation}
Then the value of $\lambda ^{*}$ is chosen to satisfy
\begin{equation}
D( p_{\lambda^{*}} || p_{A} )=D( p_{\lambda^{*}} || p_{B} )
\end{equation}

\end{lemma}

\begin{figure}[!h]
\centering
\subfigure[] { \label{Fig:px}
\includegraphics[width=0.7\columnwidth]{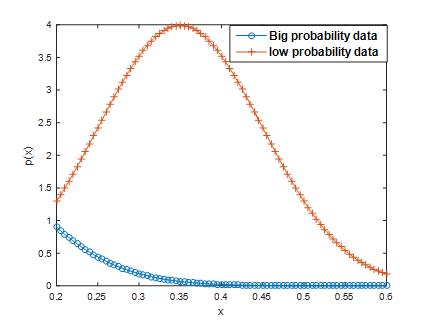}}
\subfigure[]{ \label{Fig:wpx}
\includegraphics[width=0.7\columnwidth]{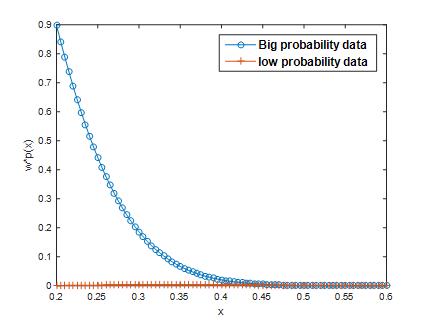}}\\
\caption{Sub-figures (a) shows the conditional probability distributions of observed data denoted as $p_{A}(x)$ and $p_{B}(x)$. $p_{A}(x)$ is $N(0,0.126^{2})$ and $p_{B}(x)$ is $N(0.35,0.1^{2})$. (b) shows $w_{A}p_{A}(x)$ ($w_{A}$=0.999) and $w_{B}p_{B}(x)$ ($w_{B}$=0.001) which are referred to by the Bayesian detection. Both graphs depict the range $[0.2,0.6]$ which is near the center of the small probability event.} \label{Fig:p}
\end{figure}

Lemma 1 (Chernoff) gives us the estimated overall error $P_{e}$ of the Bayesian detection. However, this is only the average error with respect to the miss testing rate and false alarming rate as defined in (\ref{Pe}). Note that the prior probability $w_{B}$ can be much smaller than $w_{A}$ in the detection of minor probability events, $P_{e}$ can still be small even if the miss testing rate $\alpha$ is very large. However, as mentioned before, $\alpha$ is actually the main concern of such problem. Therefore, it is necessary to make some adjustments to the traditional Bayesian detection rather than applying it directly.

The high miss testing rate of the Bayesian detection for minor probability events can be further explained by Fig .\ref{Fig:p}. In Fig .\ref{Fig:px}, $p_{A}(x)$ is $N(0,0.126^{2})$ and $p_{B}(x)$ is $N(0.35,0.1^{2})$. In Fig .\ref{Fig:wpx}, the prior probabilities are $w_{A}=0.999$ and $w_{B}=0.001$. It is obvious that $p_{A}(x)$ is very small near the center of the minor probability event $B$. Then it is reasonable to judge values in this range as the event $B$. However, in the considered problem, the prior probability of event $B$ is actually much smaller than that of event $A$. Then as shown in Fig .\ref{Fig:wpx}, the value of $w_{A}p_{A}(x)$ is still larger than $w_{B}p_{B}(x)$, which may cause the high miss testing rate of event $B$.

To overcome such a problem, the effect of the minor probability event $B$ should be magnified. To be specific, the terms $w_{A}p_{A}(x)$ and $w_{B}p_{B}(x)$ should not be applied to the judgement directly. They should be first handled by a function $f(.)$ so that the minor probability $w_{B}p_{B}(x)$ can be magnified. In \cite{fan2016message}, a measurement of message importance is proposed which focuses on the minor probability events. The measure can magnify the effect of minor probability events so that they can be dig out. In this paper, it is chosen as the magnifier to lower down the miss testing rate.

\section{Review of Message importance measure}
In \cite{fan2016message}, the message importance measure was defined. In this section, the main definitions and results are reviewed. The properties of this function which is applied to magnify the minor probability events are explained.
\subsection{Definitions of the measure}

\newtheorem{Definition}{Definition}
\begin{Definition}
For a given probability distribution $p=(p_{1}, p_{2}, ......, p_{n})$ of finite alphabet, the message importance measure (MIM) with parameter $w$ is defined as
\begin{equation}
L(p,w)={\rm log} \sum_{i=1}^{n}p_{i}{\rm exp}(w(1-p_{i}))
\end{equation}
where $w \geq 0$ is the importance coefficient.
\end{Definition}
\newtheorem{Remark}{Remark}
\begin{Remark}
When the probability distribution $p$ contains some elements with minor probabilities, $L(p,w)$ can be very large. Then this measurement can help to dig minor probability sets. Note that the larger $w$ is, the larger contribution to the MIM a small probability event has. Thus, to manifest the importance of those small probability events, $w$ is often chosen to be quite large.
\end{Remark}

Note that the MIM in Definition 1 is actually the logarithm of the mean value of function $f(x)=x{\rm exp}(w(1-x))$
for $w>0$. For discrete probability $p$, there is $0\leq x \leq 1$. For continuous situation, $p(x)$ represents the density of distribution which can be larger than $1$.

Then the function $f(x)=x{\rm exp}(w(1-x))$ can be applied here to magnify the minor probability events. Note that the function $g(x)=x{\rm exp}(-wx)$ is actually playing the same role as the former function. The only difference between them is $\frac{f(x)}{g(x)}={\rm exp}(w)$. Then the definition of the minor probability magnifier comes up as the following.
\begin{Definition}
Given an event with the probability $p$, the message importance of this event is
\begin{equation}
{\rm MIM}(p)=p e^{-wp}
\end{equation}
Considering the continuous random variable with the distribution density $p(x)$ at $x$, the message importance at $x$ is
\begin{equation}\label{MIM}
{\rm MIM}(p(x))=p(x)e^{-wp(x)}
\end{equation}
There is $w>0$ for both definitions and $w$ is typically a large number.
\end{Definition}
 \subsection{Properties of MIM magnifier}
 For two probabilities $p(x_{1})$ and $p(x_{2})$, supposing they have the same MIM, there should be
 \begin{equation}\label{Mim:equa1}
 p(x_{1})e^{-wp(x_{1})}=p(x_{2})e^{-wp(x_{2})}
 \end{equation}

 To derive the relationship of $p(x_{1})$ and $p(x_{2})$ in (\ref{Mim:equa1}), the monotonicity properties of the MIM function (\ref{Mim:equa2}) should be discussed.
 \begin{equation}\label{Mim:equa2}
 f(p)=p e^{-wp} \,\,\,(p\geq 0, w \geq 0)
 \end{equation}
 The derivative of the function is
 \begin{equation}\label{Mim:equa3}
 \frac{df}{dp}=(1-wp)e^{-wp}\,\,\, (p\geq 0)
 \end{equation}
 From (\ref{Mim:equa3}), $f(p)$ increases as $p$ increases from $0$ to $\frac{1}{w}$. In contrast, when $p > \frac{1}{w}$, $f(p)$ decreases as $p$ increases. $f(p)>0$ holds for $p \geq 0$. When the density of distribution $p$ is extremely large, there is
 \begin{equation}
 \underset{p \rightarrow \infty }{{\rm lim} }f(p)=0
 \end{equation}

 When $p(x_{1})=p(x_{2})$, equation (\ref{Mim:equa1}) can be satisfied. However, this is not relevant to the probability magnifier. According to the properties of $f(p)$ discussed above, there should be another couple of solutions. From (\ref{Mim:equa3}), when $0 \leq  p(x_{1}) <  \frac{1}{w}$, there exists $p(x_{2}) > \frac{1}{w}$ so that ${\rm MIM}(p(x_{2})) = {\rm MIM}(p(x_{1}))$. Then $p(x_{1})$ and $p(x_{2})$ are a couple of solutions with different values. For $p(x_{2}) > p(x_{1})$, the function $f(p)$ should serve to map $p(x_{1})$ to $p(x_{2})$, which is actually a probability magnifier.

 To analyze the magnifying properties of such a function, the magnifying ratio is defined as
 \begin{equation}\label{jumpratio}
 q=\frac{p(x_{2})}{p(x_{1})}-1
 \end{equation}
 where $p(x_{2})$ is larger than $p(x_{1})$ and there is $q>0$.
 By setting $p(x_{1})=p$ and $p(x_{2})=p(1+q)$, there is
 \begin{equation}
 pe^{-wp}=p(1+q)e^{-wp(1+q)}
 \end{equation}
By solving the equation, there is
\begin{equation}\label{p=f(q)}
p=\frac{{\rm ln}(1+q)}{wq}
\end{equation}
Let $p=g(q)$, the function $g(q)$ is given by
\begin{equation}
g(q)=\frac{{\rm ln}(1+q)}{wq}
\end{equation}
The first order derivative of $g(q)$ is
\begin{equation}\label{dg}
\frac{{\rm d}g}{{\rm d}q}=\frac{\frac{q}{1+q}-{\rm ln}(1+q)}{q^{2}}
\end{equation}
Then a new function is set to be
\begin{equation}
h(q)=\frac{q}{1+q}-{\rm ln}(1+q)
\end{equation}
The derivative of $h(q)$ is
\begin{equation}
\frac{{\rm d}h}{{\rm d}q}=\frac{-q}{(1+q)^2}
\end{equation}
For $q>0$, $h(q)$ is actually a monotone decreasing function. Note that $h(0)=0$, there should be $h(q)<0$ for $q>0$. Then from (\ref{dg}), there is
\begin{equation}
\frac{{\rm d}g}{{\rm d}q}<0 \,\,\,\,(q>0)
\end{equation}
In this way, it is proved that $p$ monotonically decreases with respect to the magnifying ratio $q$ and MIM parameter $w$. Therefore, a larger $q$ is corresponding to a smaller $p$. Therefore, the smaller the initial probability $p$ is, the larger magnifying ratio $q$ it can get from the magnifier. Furthermore, when $w$ increases, $p$ is also smaller while $q$ does not change in the process. Then the magnifying effect for small probabilities can be better when $w$ is large.

\begin{figure}[tbp]
  \centering
  \includegraphics[width=1\columnwidth]{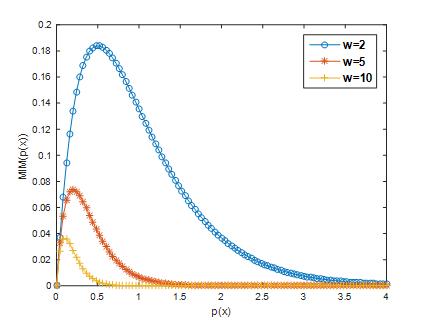}\\
  \caption{The magnifying function ${\rm MIM}(p(x))=p(x)e^{-wp(x)}$. $p(x)$ is the distribution density. $w$ for different curves are separately $2$, $5$ and $10$.}\label{Fig:MIMf}
\end{figure}

The function ${\rm MIM}(p(x))$ defined in (\ref{MIM}) is depicted in Fig .\ref{Fig:MIMf} with respect to $w=2$, $w=5$ and $w=10$. When $w$ gets larger, the probability has to be smaller to get the same magnifying ratio as that when $w$ is small. Then the probabilities which is not small enough will not be magnified.

\section{New detection approach}
In this section, based on the message importance measure, the specific criterion for detection of minor probability events is introduced. Compared with traditional Bayesian detection, it magnifies the minor probabilities so that they have larger chance to be detected. In this way, it can have better performance in detection of the minor probability events.
\subsection{Judging criterion}
\newtheorem*{Detection}{Detection Approach}
\begin{Detection}
Given two events denoted as $A$ and $B$ with prior probabilities $w_{A}$ and $w_{B}$, the conditional distribution of observed data is $p_{A}(x)$ and $p_{B}(x)$. If $w_{A}>>w_{B}$ and the magnifying function is $f(p)=p{\rm exp}(-s_{0}p)$, the modified judging criterion is
\begin{equation}
\{x_{i}|x_{i}\in A\}=\left \{ x_{i}| \frac{f(w_{A}p_{A}(x_{i}))}{f(w_{B}p_{B}(x_{i}))}<1 \,\, {\rm and} \,\, \frac{w_{A}p_{A}(x_{i})}{w_{B}p_{B}(x_{i})}>1 \right \}
\end{equation}
\begin{equation}
\{x_{i}|x_{i}\in B\}=\left \{ x_{i}|\frac{f(w_{A}p_{A}(x_{i}))}{f(w_{B}p_{B}(x_{i}))}>1 \,\, {\rm or} \,\, \frac{w_{A}p_{A}(x_{i})}{w_{B}p_{B}(x_{i})}<1 \right \}
\end{equation}
\end{Detection}
\begin{Remark}
The choice of $s_{0}$ should come from the training data which represents the experience. It need not necessarily contain samples with minor probability events which is not easy to obtain. Then given the data $x$ of normal events $A$, when the $w_{B}p_{B}(x)$ is projected by the magnifier from range $[0,\frac{1}{s_{0}}]$ to range $[\frac{1}{s_{0}},\infty ]$, it should still be smaller than $w_{A}p_{A}(x)$. In the following simulation part, given samples $X$ from distribution $p_{A}(x)$, the value of $s_{0}$ satisfies $f({\rm Mean}(w_{A}p_{A}(X)))=f({\rm Mean}(w_{B}p_{B}(X)))$, where ${\rm Mean}(.)$ represents the average value of the term with respect to the samples.
\end{Remark}

\subsection{Link with traditional Bayesian detection}
Given the observed data $x$, the judgement $p(Q|x)$ can be calculated by the following Bayesian equation
\begin{equation}\label{Bayes}
p(Q|x)=\frac{p(x|Q)p(Q)}{p(x)}
\end{equation}
where $Q$ represents the event under judgement.

By comparing $p(Q|x)$ of the events $A$ and $B$, the judgement can be obtained. For $p(x)$ is the same for both events, the judgement can be done by directly comparing $w_{A}p_{A}(x)$ and $w_{B}p_{B}(x)$.

When magnifying the minor probability events by the message importance measure, the judgement criterion should be ${\rm MIM}(p(Q|x))$. Then from (\ref{MIM}) and (\ref{Bayes}), there is
\begin{equation}
{\rm MIM}(p(Q|x))=\frac{p_{Q}(x)w_{Q}}{p(x)}{\rm exp} ( -s_{0}\frac{p_{Q}(x)w_{Q}}{p(x)} )
\end{equation}
For $p(x)$ is not relevant with the event $Q$, it can be omitted when comparing ${\rm MIM}(p(Q|x))$ of events under judgement. Therefore, the criterion of comparing $p_{Q}(x)w_{Q} {\rm exp}(-s_{0}p_{Q}(x)w_{Q})$ in the detection approach can be derived.

As shown in Fig .\ref{Fig:MIMf}, the function ${\rm MIM}(p(x))$ increases for $p(x)$ in $[0,\frac{1}{s_{0}}]$ and decreases for $p(x)$ in $[\frac{1}{s_{0}}, \infty]$. Then tiny element $w_{B}p_{B}(x)$ in $[0,\frac{1}{s_{0}}]$ can have the same MIM value as a point $(w_{B}p_{B}(x))^{'}$ in $[\frac{1}{s_{0}}, \infty]$. In this way, the tiny element $w_{B}p_{B}(x)$ can be magnified to $(w_{B}p_{B}(x))^{'}$ and compared with $w_{A}p_{A}(x)$. Therefore, the miss testing rate can be largely reduced.

\section{Simulation results}
In this section, the detection strategy is tested on two Gaussian distributions. The miss testing rate and false alarming rate of the proposed strategy and traditional Bayesian method are separately recorded. It shows that when prior probability of the minor probability event keeps decreasing, the proposed method can obtain better performance.

\begin{figure}[tbp]
  \centering
  \includegraphics[width=1\columnwidth]{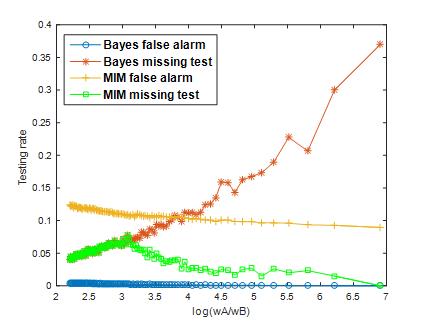}\\
  \caption{The miss testing rate and false alarming rate of both detection strategies with respect to the log ratio between the two prior probabilities. The conditional distributions $p_{A}(x)$ and $p_{B}(x)$ are both Gaussian distributions. }\label{Fig:c_W}
\end{figure}

In Fig .\ref{Fig:c_W}, the miss testing rate and false alarming rate of Bayesian detection and MIM based detection are depicted. The conditional distributions $p_{A}(x)$ is $N(0,0.126^{2})$ and $p_{B}(x)$ is $N(0.5,0.1^{2})$. The parameter $s_{0}$ is selected by the training data from $p_{A}(x)$. In this case, $w_{B}$ is the minor prior probability and $w_{A}$ is the prior probability of the normal event. There is
\begin{equation}
w_{A}+w_{B}=1
\end{equation}
The $x$-axis in Fig .\ref{Fig:c_W} represents ${\rm log}(\frac{w_{A}}{w_{B}})$.

As shown in Fig .\ref{Fig:c_W}, when the minor prior probability $w_{B}$ keeps decreasing, the MIM based method can have decreasing miss testing rate while maintain a stable and acceptable false alarming rate. However, the conventional Bayesian method has high miss testing rate in this case.

\begin{figure}[tbp]
  \centering
  \includegraphics[width=1\columnwidth]{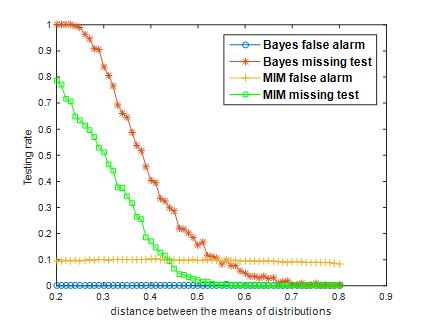}\\
  \caption{The miss testing rate and false alarming rate of both detection strategies with respect to the distance between $E(p_{A}(x))$ and $E(p_{B}(x))$. The conditional distributions $p_{A}(x)$ and $p_{B}(x)$ are both Gaussian distributions.}\label{Fig:c_mu}
\end{figure}

In Fig. \ref{Fig:c_mu}, $p_{A}(x)$ is $N(0,0.126^{2})$ and the variance of $p_{B}(x)$ is still $0.1^{2}$. ${\rm E}(p_{B}(x))$ changes from $0.2$ to $0.8$. $w_{A}$ is $0.992$ and $w_{B}$ is $0.008$. The x-axis of the figure represents the distance of the means of $p_{A}(x)$ and $p_{B}(x)$. From the graph, it can be seen that the MIM based method has better miss testing rate compared with the conventional Bayesian method. The conventional Bayesian method only gains equal testing quality when the means are far enough. Besides, the false alarming rate of MIM-based method is also stable and acceptable in this process.

\section{Conclusion}
In this paper, a new approach to detect the minor probability events in big data was proposed. Based on the message importance measure, the minor probabilities is magnified so that they can be detected more easily. By simulations, it was verified that the MIM based detection method could have much lower miss testing rate while maintaining an acceptable false alarming rate. This advantage can meet the needs of minor probability events detection in big data. The method can make up for the traditional Bayesian method and help to dig abnormal events in big data.

\ifCLASSOPTIONcaptionsoff
  \newpage
\fi

\end{document}